\documentclass[12pt]{article}

 \newread\testifexists
 \def\GetIfExists #1 {\immediate\openin\testifexists=#1
     \ifeof\testifexists\immediate\closein\testifexists\else
     \immediate\closein\testifexists\input #1\fi}

 \usepackage{amsfonts}
 \usepackage{amssymb}
 \usepackage{graphicx} \usepackage{epstopdf}
 \immediate\openin\testifexists=e
     \ifeof\testifexists\immediate\closein\testifexists\else
     \immediate\closein\testifexists\input e\fipsf

 \def\Bbb#1{\setbox0=\hbox{$\tt #1$}  \copy0\kern-\wd0\kern .1em\copy0}

 \def\bbf#1{\setbox0=\hbox{$#1$} \kern-.025em\copy0\kern-\wd0
         \kern.05em\copy0\kern-\wd0 \kern-.025em\raise.0433em\box0}

 \immediate\openin\testifexists=a
     \ifeof\testifexists\immediate\closein\testifexists\else
     \immediate\closein\testifexists\input a\fimssym.def  

                     \newcommand{\fn}{\footnote}
              
 \newcommand{\be}{\begin{eqnarray}}             \newcommand{\ee}{\end{eqnarray}}
 \newcommand{\bi}[1]{\begin{itemize}\item[#1]}           \newcommand{\ei}{\end{itemize}}

 \newcommand{\crlb}[1]{\label{#1}\\[2pt]}
 \newcommand{\eela}[1]{\,\hbox{\scriptsize{#1}\qquad}\label{#1}\end{eqnarray}}
 \newcommand{\eelb}[1]{\label{#1}\end{eqnarray}}
 
 \newcommand{\newsecb}[2]{\section{#1}\label{#2}\setcounter{equation}{0}}
 	 	
	 \newcommand{\subsecb}[2]{\subsection{#1}\label{#2}}
	 	 	
	 	 \newcommand{\subsubsecb}[2]{\subsubsection{#1}\label{#2}}

 \newcommand{\nolabels} {\def\eel{\eelb} \def\crl{\crlb} \def\newsecl{\newsecb}\def\subsecl{\subsecb}
 	\def\subsubsecl{\subsubsecb}\def\bibiteml{\bibitem}\def\citel{\cite}\def\labell{\label}}

\newcommand\publishversion{\nolabels\setlength{\textheight}{9in}\setlength{\oddsidemargin}{0in}
    \setlength{\textwidth}{6.3in}\setlength{\topmargin}{-.7in}}

 \def\a{\alpha}               
 \def\d{\delta}           
             \def\m{\mu}
 \def\f{\phi}        \def\F{\Phi}        
 \def\j{\psi}                   \def\SS{\Sigma}

 \def\iss{\ =\ }
 
 \def\th{\(^{\!\mathrm{th}}\)}
 \newcommand{\rem}[1]{}	

 \def\ffract#1#2{\raise .2 em\hbox{$\scriptstyle#1$}\kern-.35em/
                 \kern-.2em\lower .15 em \hbox{$\scriptstyle#2$}}
 
   \def\halff{\ffract1{\!2}}
 
   \def\low#1{\raise-.3em\hbox{$\scriptstyle{#1}$}} 
 	\def\lowtje#1{\raise-.2em\hbox{$\!\scriptscriptstyle{#1}\!$}}
  
\def\bmatrix{\begin{matrix}} \def\ematrix{\end{matrix}} \def\bpmatrix{\begin{pmatrix}}\def\epmatrix{\end{pmatrix}}
\def\bcenter{\begin{center}} \def\ecenter{\end{center}}

\def\lowerheightfig#1#2#3{\(\raise-#1\hbox{\includegraphics[height=#2]{#3}}\)}
\def\lowerwidthfig#1#2#3{\(\raise-#1\hbox{\includegraphics[width=#2]{#3}}\)}

\def\glt{\hbox{\,\raise .35em\hbox{$>$}\kern-.8em\raise-.15em\hbox{$<$}\,}}  
 \def\ol{\overline}

\publishversion
\begin{document} \begin{titlepage}

\title{
\vskip 20mm \LARGE \bf Nature's Book Keeping System\fn{A contribution to the collection ``Space, Time, and Frontiers of Human Understanding".}}
\author{Gerard 't~Hooft}
\date{\normalsize Institute for Theoretical Physics \\
\(\mathrm{EMME}\F\)   \\
Centre for Extreme Matter and Emergent Phenomena\\[5pt] 
Science Faculty\\ 
Utrecht University\\[5pt]
 POBox 80.195 \\
 3808TD, Utrecht  \\
The Netherlands  \\[10pt] 
e-mail:  g.thooft@uu.nl 
\\ internet:  http://www.staff.science.uu.nl/\~{}hooft101/ }
\maketitle
\abstract{Establishing how one should describe and study nature's fundamental degrees of freedom is a notoriously difficult problem. It is tempting to assume that the number of bits (or qubits) needed in a given Planckian 3-volume, or perhaps 2-volume, is a fixed finite number, but this ansatz does not make the problem much easier. We come not even close to solving this problem, but we propose various ingredients in phrasing the questions, possibilities and limitations that may serve as starting points.}
\vfill \flushleft{Version April, 2016}\\
{\footnotesize Last typeset: \today.}
\end{titlepage}

 \setcounter{page}{2}


\newsecl{Introduction}{intro} 
When formulating a theory concerning the physical world, one usually just assumes that the basic kinematical variables are known, and can be enumerated, and studied by some experimental technique. Think of stars and planets in some star system, a star or planet itself, or eventually some agglomeration of elementary particles, as the case may be. The more basic and interesting part of the theory is then the attempt to explain the behaviour of these variables, in terms of some mathematical scheme, often using symmetries to make the problem more manageable.

One clearly assumes then that we know what the main variables are, how their properties, such as coordinates and momenta, can be listed, and most importantly, how we can maintain a book keeping system such that the results of our calculations can be checked against experimental observations. 

In many practical cases, such a book keeping system appears to be totally under control, so that it requires little further discussion. Assuming that natural phenomena have a deductive ordering, we are tempted first to focus on nature's book keeping system at the tiniest distance scales. Once that it understood, we can go to higher levels and arrange our schemes for those branches of physics as well.

In this treatise,  we focus on the more primary question: what \emph{are} nature's most fundamental kinematic variables, and how do we find out about them? In short:  what is nature's book keeping system at the most fundamental level?

\newsecl{The sub-atomic world}{subatom}

Today, the terrain of the tiniest distance scales of physics where things are still pretty clear, is the world of the sub-atomic particles. How to do the book keeping system there, and indeed also how to perform the next step, which is the detailed description of nature's evolution laws, was unravelled during most of the 20\th century. 

The  way to list the dynamical degrees of freedom for the subatomic particles is dictated by nature's symmetry laws, notably special relativity, in combination with the laws of quantum mechanics. We are in the fortunate situation that this combination is so restrictive that only one scheme survived, called quantum field theory. The situation still seemed to be so complex that most of the 20\th century was needed to handle all obstacles, but the outcome was crystal clear.

At first sight, there appear to be two distinct descriptions for the subatomic particles, to be referred to as \emph{field} space and \emph{Fock} space.
When we choose the field space option, we register the functional distribution of all \emph{fields} in the system, of which there are three types: the gauge connection fields \(A_\m^a(\vec x,t)\), which are vectors in 4 dimensional Minkowski space-time, the spinorial Dirac fields \(\j_\a^i(\vec x,t),\ \ol\j^{\,\a}_i(\vec x,t)\), which form a Grassmann algebra, and some scalar fields \(\f^a(\vec x,t)\). Now by far the most practical way to register the fields is to list the values of their Fourier components, so that the fields are expanded in terms of plane waves. Each of these waves behaves as a harmonic oscillator. One subsequently subjects these oscillators  to the rules of quantum mechanics. In accordance with whatever is dictated by the (relativistic) field equations, the oscillators will be weakly or strongly coupled. The (multi-dimensional) Schr\"odinger equations resulting from this can now be handled by standard methods in mathematical physics.

The Fock space solution may seem to be something altogether different. Here, we do not talk of fields, but of states with arbitrary numbers of elementary particles present. These particles, and their anti-particles, may have spin 1, \(\halff\), or 0, and we list which of them are present, and what their coordinates \(\vec x\) are, at the time \(t\). The particles may form quantum wave functions, and their motion is dictated by relativistic equations. The particles may interact weakly or strongly, and their energies can be calculated. During an interaction, particles can be created or destroyed.

Surprisingly,  it was found that these two ways to register what is going on, are different sides of the same coin. The operator fields describing the creation or annihilation of particles, are to be identified with the fields mentioned previously, while the energy quanta of these fields are identified with the particles. One then discovers that both descriptions are totally equivalent. They lead to exactly the same predictions for the outcomes of an experiment. Analysing the situation further, one finds that both schemes are subject to the \emph{energy conservation law}: every configuration has a characteristic energy, and the total energy is a positive, conserved quantity. This law dictates that simple field or particle configurations also evolve into simple configurations. More involved configurations require more energy. Thus, we should be aware of the fact that the field\,/\,Fock space method owes its success to the energy conservation law.

\newsecl{Gravitation}{grav} 

 \def\ffract#1#2{\raise .2 em\hbox{$\scriptstyle#1$}\kern-.3em/
                 \kern-.3em\lower .15 em \hbox{$\scriptstyle#2$}}

All this is not at all the final word on nature's book keeping system. What was kept out of the picture so-far, was the gravitational force. This force obeys Einstein's equations for gravity, but these are subject to a new symmetry: general relativity.

Of course, what physicists in the second half of the 20\th century tried to do, is to adapt the field\,/\,Fock description of nature's degrees of freedom to these new field equations, discovering that also new particles are needed: \emph{gravitons}, which carry spin 2, and perhaps \emph{gravitinos}, with spin \(\ffract 32\). What went wrong?

Technically, at least two things went wrong. One was that, space and time, or the coordinates that ought to be used to register our particles and our waves, themselves become dynamical, due to space-time curvature, as if the books in which we want to write the whereabouts of our particles, began to lead lives of their own.
Closely related to this was the problem that the new field equations, and the particle equations alike, became \emph{non-renormalizable}. This means that the ultra-small distance behaviour runs out of control. 

But something else went wrong: something wicked happens with the energy conservation law: gravity is an \emph{attractive} force, and it becomes stronger if mass, or equivalently, energy, conglomerates. Energy increases if we concentrate on smaller distance scales. The energies of particles, regardless how much mass they have, can approach to zero and even, in a sense, become negative. The strength of the gravitational force then totally runs out of control. 

An other way of saying this is as follows. Since gravity \emph{attracts} equal sign masses, while electric and magnetic forces, as well as the weak force, act repulsively when charges carry equal signs, gravitational fields carry a negative field energy density, which may compensate for any positive amount of energy that other particles or fields might have. This implies that the energy conservation law can no longer be called upon to keep the laws of physical interactions under control. Nature's book keeping rules will here have to be \emph{entirely} different from what we are used to in particle physics.

In fact, there seems to be no strong impediment against the formation of ultra-tiny black holes. If two particles approach one another too closely with too much energy, black hole formation is inevitable. 
In quantised general relativity, neither Fock space nor field space provide useful lists of all possible configurations. It should not be a surprise that we still have not come with credible, let alone complete, laws on how things evolve at such small distances. The distance scale at which things are sure to go haywire, is the so-called Planck length, 
	\be L_{\mathrm{Planck}}\iss\sqrt{\hbar\, G/c^3}\iss 1.6162\times 10^{-33}\,\mathrm{cm}\ , \ee
where \(G\) is Newton's constant. This is roughly a billion times a billion times smaller than the sizes of particles studied in the Large Hadron Collider. Now imagine a volume of space of just a few Planck units cubed. What can we suggest about nature's book keeping system to describe \emph{anything} one \emph{can} imagine sitting in such a tiny region?

\newsecl{Superstrings and black holes}{sustr}

There are numerous clues that may be considered. A very powerful proposal is that of \emph{superstring theory}. Representing the tiniest objects in physics as being line-like instead of point-like, yielded equations that are much more restrictive than the older relativistic equations for point-like objects, and the renormalization difficulties seem to disappear. Strings refuse to be crammed into tiny volumes because of their symmetry constraints. 

There is much support for this idea, but I do see various problems with it. One is that, just because strings cannot be folded into tiny volumes, the notion of locality seems to be in jeopardy. An other is the lack of a local energy law, which could destabilise these systems. Then, even if infinities may cancel out, there still seem to be no obvious bounds as to how much information can sit in a tiny volume. Finally, there is the issue with the microscopic black holes. 

In short, strings seem to be far from orderly book keepers. We should search for something better. Of course, we do not exclude the possibility that, after finding a more satisfactory answer, string theory may return to become the most promising game in town.

A more systematic attack comes from the black holes themselves. Black holes as tiny as the Planck scale will be difficult to understand as yet, but when their typical length scale, the radius \(r=2GM/c^2\), is more than an order of magnitude bigger,
then standard quantum field theory dictates their behaviour: such black holes emit all sorts of particles, with a an apparently perfectly thermal\fn{This neat picture has recently been questioned; thermodynamics alone does not suffice for the book keeping of particles near a black hole horizon\,\cite{GtHentanglement}} spectrum, called \emph{Hawking radiation}\,\cite{hawkingrad}. From this calculation, it is easy to derive the entropy of a black hole, and a little statistical physics then reveals a startling feature: \emph{The total number of distinct quantum states a black hole can be in, is an exponential function of the horizon area.}\,\cite{bhmicrostates}

This is only a small number of states, and the fact that it depends on the area rather than the total volume, came as a surprise. It looks as if black holes require a \emph{very simple} book keeping law: one boolean degree of freedom (something like a spin variable that can only take the values \(\pm 1\)), for every fundamental little square on the horizon. The fundamental surface of such a square will be 
	\be \d \SS=(4\,\log 2)\,L_{\mathrm{Planck}}^2\ . \ee
Since black holes seem to be the most compact form of matter that can exist in nature, having one boolean degree of freedom per unit of surface area seems to be an absolute maximum.\cite{bekenstein} This is referred to as the \emph{holographic principle}: nature's information content can be measured in terms of bits per \(cm^2\), with the inverse Planck length squared as an absolute bound.\cite{holo} One gets the impression that nature might obey a beautiful book keeping system. 

All we have to do now is insert the details, and figure out nature's laws about processing these data.

This turns out to be prohibitively difficult today. One difficulty is Lorentz invariance. Suppose we have a situation with the maximal amount of information on a large surface. Now, perform a Lorentz transformation that keeps the surface in place. We get an other configuration, which is the previous one, but Lorentz contracted. There seems to be no limit on how many Lorentz contractions one can perform, but there should be such a limit, otherwise our book keeping system would not be finite.

Black holes do suggest what to do here: according to the statistical interpretation of the Hawking radiation emitted by the black hole, particles going in transfer their information onto the particles going out. One can calculate how this happens, by studying the curvature of space-time as caused by the gravitational fields of the in- and out-going objects.\cite{GtHblinfo} The longitudinal location of the out-particles turns out to depend on the momenta of the in-going ones, and this is an important clue: nature's book keeper throws in-going particles into the basket of out-going ones, if only we could understand in more detail how this happens. We  \emph{calculate} how the gravitational force between in-going particles and out-going particles brings about this behaviour. These investigations may pay off already at the domain of the Standard Model itself: all particles species will have to leave their unique imprints in the fabric of space and time. Yet, this calculation did not lead to a properly finite book keeping device -- although we think we came close.\,\cite{GtHentanglement}\cite{GtHbhentropy} 

In short, the difficulty seems to be that Einstein's gravity equations include fields that describe the curvature of space and time. This adds a twist to our story: nature has to write its book keeping records on crumbled material: curved space-time. How do we make sense of that?

A daring answer could be that, for putting down information regarding space, time and matter, one has to replace this folded space-time material by something that is almost flat. It would be against Einstein's general invariance principle, but we may have to prepare for prices of this kind that will have to be paid. Nature might be giving us clues concerning this point: our universe appears to be \emph{almost} flat, too flat to make sense in an uncensored Einsteinian theory.

One guiding principle was strongly adhered to by practically all researchers: anything as small as molecules, atoms, subatomic particles and other energy quanta,  or smaller, always required the language of \emph{quantum mechanics} to describe their properties. Most researchers take this to mean that we will need to use this quantum mechanical language no matter which approach we try. Yet, here also, one can have doubts. To me, quantum mechanics seems to be a \emph{tool} rather than a theory. This means that, the most commonly employed interpretation of quantum mechanics, a discipline that sometimes seems to challenge our sense of logic to the extreme, may not always be justified. One can speculate that nature's ultimate book keeping device at the tiniest possible scales, must be simpler than any quantum mechanical one\,\cite{GtHqm}.

One should immediately add that, giving up general relativity and/or quantum mechanics, will bestow on us the burden to explain why both principles work so well in all domains of physics that have been explored at present.

Numerous other schemes have been proposed, over time. As is often the case in situations such as these, there have been an abundance of completely wild concoctions that serious researchers have come up with. Lack of phantasy is not our problem\,\cite{sorkin}\cite{rovellismolin}\cite{ambjornloll}. The question is rather, how to find a direct chain of solid arguments, with many calculations on the way to corroborate intermediate conclusions, leading to an unavoidable result that explains how particles, fields, energy quanta or \emph{whatever}, can be listed in any tiny domain of our physical world, such that, when we take into account the physical laws controlling their behaviour, we find out how things behave at larger scales, until we hit the domain of ordinary quantum field theories, with which we can calculate almost anything we wish to know. 

Note our emphasis on scales and sizes of things. In ordinary physics there seems to be no limit on how small one can imagine fundamental units of matter to be. In an orderly controlled physical world, there must come an end to these scale transformations. There is one natural way in which one may imagine such an ``end of scales" to arrive in our book keeping system: scaling symmetry or more precisely, the introduction of \emph{exact} local scale- and conformal invariance. One can then employ the \emph{same} book keeping variables to describe all scales to come. The use of such a symmetry, which must be a \emph{local} symmetry, is quite well-known in theoretical particle physics, and is called local gauge invariance. Local gauge invariance can be `spontaneously broken' (the well-known BEH mechanism). But the known gauge theories mostly refer to internal rotation groups, not scaling transformations. Again, here General Relativity may help us to explore new grounds: coordinate transformations do include scale transformations, and this means that a locally conformal invariant description of gravitational effects\,\cite{Mannheim}\cite{GtHconfgrav} is possible, in principle. It is the present author's view that one must look at gravitation from this perspective. General relativity features an exact local conformal gauge symmetry, which is spontaneously broken just as in BEH. 

Thus, we emphasise, the bookkeeping system that we are after, should be able to describe literally all states that can occur in the physical world.
Without going into any details, let us briefly recapitulate what we may have arrived at today.

\newsecl{Nature's ultimate book keeping system}{ultimate}

When trying to describe nature's most general state, we have to ask what the ultimate state will look like when we squeeze as much matter as is possible into the tiniest possible volume. One usually expects one or more black holes to form. All presently known laws of fundamental physics make use of quantum mechanics, so, even if one is skeptical about this, it is natural first to search within the quantum formalism. A very important notion then is \emph{unitarity}: two states that start out to be orthogonal and normalised, evolve into another pair of states that is also orthogonal and normalised. 

Laws of gravity, as they are known today, then suggest that all forms of matter will be \emph{geometric}: the way they affect the curvature of space and time is the only form of information that is conserved\,\cite{GtHentanglement}--\cite{GtHbhentropy}. We think we drew the important conclusion that observations of the sort mentioned here, will be the only way to reconcile finiteness of the degrees of freedom with an unbounded group of local Lorentz transformations. This is extremely important, if true. It means that Fock space will not be the appropriate language; rather, we get something that resembles a bit more string theory, which is also basically geometric. String theories known today, however, seem not yet to be based on very sound book keeping. 

 Indeed, for this, and other reasons, string theory has been under attack\,\cite{Smolin}. According to Smolin, the internal logic in string theories is too hazy to serve as a sound physical prescription. Consequently, it is all but impossible to extract reliable predictions from string theories, other than the claim that everything will soon be understood. These attacks are understandable, but not totally justified. String theory is a collection of quite impressive mathematical constructions and theorems. The feeling, shared by many of its practitioners, is that these mathematical notions must mean something, and that our physical world is likely to require such ingredients. String theory tells us that there are not only point particles, described by fields that can be used for a Fock space formalism, but in addition, we have stringlike features, as well as two-dimensional membranes and higher dimensional structures. This is an important lesson, and it may well be that this idea will be important for our theoretical thinking in the future.
 
 The demand from string theory that space and time themselves must feature either   10 or   26 dimensions, however, seems to be too restrictive. If indeed, as we suspect, physical degrees of freedom form discrete sets, then dimensionality of space and time may be less well-defined notions, so that such `predictions'  from string theory may well be due to some mathematical idealisation having little to do with reality. All in all, we are badly in need of a more orderly listing of all conceivable configurations of physical variables in a small region of space and time. 
 
 As the reader may notice, this demand by itself is also not formulated very sharply.  This is exactly the point we wish to bring forward in this paper. Part of our problem is the precise formulation of our question or questions.  Nature's book keeping system must be outlined together with the answers to the question how the variables will evolve in time.  it is important to observe that the big revolutions in science often came with improvements in our way to phrase the physical degrees of freedom, together with new proposals as to how these evolve. The grand total is what we call physical law. 
 
 Smolin complains that science today is slowing down considerably, and blames this to the rise of string theory and its staggering promises, at the expense of other approaches in the basic sciences. 
 
 But Smolin forgets that the real revolutions in science are often only recognised in hindsight. To our judgement, it is quite conceivable that many or all of our questions concerning nature's book keeping system will be solved in the not so distant future.
 However, the road towards these solutions will consist of very small but mathematically precisely formulated steps in our way of thinking. String theory was an interesting guess, but may well have been a too wild one. We are guessing the mathematical structures that are likely to play a role in the future, but we fall short on grasping their internal physical coherence and meaning. For this, more patience is needed.

\end{document}